\documentclass[conference]{IEEEtran}

\usepackage[pdftex]{graphicx}

\usepackage{xcolor}
\usepackage{amssymb}
\usepackage{yfonts}
\usepackage{cite}[biblabel]

\begin{document}

\title{Detecting the early inspiral of a gravitational-wave signal with convolutional neural networks}

\author{\IEEEauthorblockN{{ Grégory Baltus} and \\ Jean-René Cudell}
\IEEEauthorblockA{Institut de Physique, Bâtiment B5 \\ Université de Liège\\ Sart Tilman B4000 Liège, Belgium\\
Email: gbaltus@uliege.be, jr.cudell@uliege.be}
\and
\IEEEauthorblockN{Justin Janquart, Melissa Lopez, Amit Reza\\ and  Sarah Caudill}
\IEEEauthorblockA{Nikhef Science Park 105,\\ 1098 XG, Amsterdam, The \\Netherlands.
Institute for Gravitational \\and Subatomic Physics (GRASP),\\ Utrecht University, Princetonplein 1,\\ 3584 CC, Utrecht, The Netherlands\\
Email: j.janquart@uu.nl, m.lopez@uu.nl, \\ areza@nikhef.nl, s.e.caudill@uu.nl}}
%\and

%\IEEEauthorblockN{Amit Reza}
%\IEEEauthorblockA{Department of Physics \\ Indian Institute of Technology \\ Gandhinagar, Gujarat 382355, India \\ Email: amit.reza@iitgn.ac.in}}

\maketitle

\begin{abstract}
We introduce a novel methodology for the operation of an early %warning 
alert system for gravitational waves. It is based on short convolutional neural networks. We focus on compact binary coalescences, for light, intermediate and heavy binary-neutron-star systems.
The signals are 1-dimensional time series $-$ the whitened time-strain $-$ injected in Gaussian noise built from the power-spectral density of the LIGO detectors at design sensitivity. We build short 1-dimensional convolutional neural networks to detect these types of events by training them on part of the early inspiral. We show that such networks are able to retrieve these signals from a small portion of the waveform.
\end{abstract}

\IEEEpeerreviewmaketitle

\section{Introduction}
Since 2015, the Laser Interferometer Gravitational Wave Observatory (LIGO)~\cite{TheLIGOScientific:2014jea} and Virgo~\cite{TheVirgo:2014hva} collaborations have detected multiple gravitational wave (GW) signals coming from coalescences of compact binary objects. Most of these signals come from binary black holes (BBH), but others come from binary neutron stars (BNS), such as GW170817~\cite{TheLIGOScientific:2017qsa} or GW190425~\cite{Abbott:2020uma}. The latter events are expected to have an electromagnetic counterpart, which if detected, can enable us to improve our understanding of complex astrophysical and cosmological effects in the context of Multi-Messenger Astronomy (MMA), and lead to novel tests of General Relativity \cite{cowperthwaite2017electromagnetic, sengupta2019first, fishbach2019standard, berti2018extreme, abbott2019tests}. It is thus crucial to detect GW events rapidly and send early alerts to other observational facilities so that they have enough time to point the different instruments to the location of these events and detect this electromagnetic signature as soon as it occurs.

A gravitational wave is made of three parts: a very long inspiral (when the bodies rotate around each other), increasing in amplitude, followed by a high-amplitude merger (when they touch and join), then by a ringdown (when the newly formed body returns to  its ground state). For BNSs, only the low-amplitude early inspiral and the higher-amplitude late inspiral have been detected so far. When it enters the detectability band of the 
interferometers by reaching their frequency threshold, the BNS gravitational wave is in its early inspiral region. 
This threshold will decrease as we upgrade our detectors and enrich the current network with KAGRA and LIGO-India in the upcoming years~\cite{Akutsu:2020his, Unnikrishnan:2013qwa}, and   Cosmic Explorer, the Einstein Telescope (ET), and LISA~\cite{Reitze:2019iox,Punturo:2010zz,Sathyaprakash:2012jk,Audley:2017drz} in the long run. These coming upgrades will be more sensitive at low frequency. This will increase the duration of the signals in the detectability range, and make early warning predictive. 

%%%%---------------------------------------------------------------
The standard method used to detect Compact Binary Coalescence (CBC) is \textit{matched filtering}. This method relies on template banks, and correlates these templates with the data over the sensitivity band of the detectors, allowing signals to be extracted from the detector noise. Such a technique can be computationally intensive~\cite{cannon2012toward,usman2016pycbc, dal2014implementing, 2017PhRvD..95d2001M}, particularly in the coming years as the number of detections increases and the GW signals become longer.

The templates used in the usual matched filtering for neutron stars correspond to the full length of the inspiral, but recent efforts have shown that it is possible to detect a signal with only a fraction of the waveform~\cite{Sachdev:2020lfd}. It is important to note that the signal-to-noise ratio (SNR) accumulates more slowly in the  early inspiral than it does in the late inspiral/merger. That means that it is more difficult to detect a signal with only a few seconds of early inspiral than with the same number of seconds of late inspiral/merger. 
%%%%---------------------------------------------------------------

An interesting alternative has recently emerged: Deep Learning (DL), which offers new tools in the GW field. These are able to perform analyses rapidly since all the intensive computation is diverted to the one-time training stage. This makes these methods orders of magnitude faster than conventional \textit{matched filtering} techniques~\cite{George:2016hay}. 
Due to this fact DL methods and in particular convolutional neural networks (CNNs) sparked the interest of several authors, who have built algorithms to demonstrate their power when searching for CBC  events~\cite{George:2016hay, George:2017pmj, Gabbard:2017lja, wei2020deep, krastev2020real}.  In addition, in \cite{George:2016hay} and \cite{George:2017pmj}, the authors demonstrate that neural networks are able to generalize well to unseen data sets, even when glitches are present.

%%%%---------------------------------------------------------------
In the context of early warning, a recent investigation has shown the possibility to detect GW signals presented as spectrograms by training only on the inspiral part and using a \emph{ResNet50} network~\cite{wei2020deep}. However, the creation of these spectrograms is an additional pre-processing step. We found that the computation of such a spectrogram takes around~0.5 seconds depending on the desired resolution.   
%%%%---------------------------------------------------------------

To overcome this challenge, we propose to build a 1D CNN-based pipeline that treats the whitened strain to detect the early inspiral phase of GWs in low latency, by training only on this part of the waveform. We would like to emphasize that this work is a proof of concept that demonstrates the promises of this line of research.

The signals are injected into simulated Gaussian noise, which is built from a PSD that corresponds to the design sensitivity of advanced LIGO (aLIGO)~\cite{aasi2015advanced}. Here, we consider 3 different event categories depending on the masses of the progenitors: light, intermediate and heavy BNS\footnote{The number of categories is arbitrary and could be optimized in the future. Indeed, one can choose a larger number of categories with more restricted mass ranges. However, it would require more GPUs to make the whole analysis.}. We then build  three short neural networks with a few convolutional layers, as has been previously done in the literature \cite{George:2016hay, George:2017pmj, Gabbard:2017lja}. 

The structure of this paper goes as follows: in Sec.~\ref{sec:PartaiSNR} we describe how our signals are simulated. In Sec.~\ref{sec:method} we present the methodology to train short 1D CNNs and evaluate their performance.  In Sec.~\ref{sec:Discussion}  we present our results, and in Sec.~\ref{sec:Conclusions} we draw our conclusions and outline the next steps towards a fully functional ML-based pipeline for early alerts. 
%%%%---------------------------------------------------------------

\section{Data Generation}
\label{sec:PartaiSNR}
%%%%--------------------------------------------------------------
\subsection{SNR and partial inspiral SNR}
A key element of the data analysis for GWs is the signal-to-noise ratio (SNR). It  provides a measure of the degree to which a given template matches the measured signal. 
Its mathematical expression is as follows:
%%%%---------------------------------------------------------------
\begin{equation}\label{eq:SNR}
    \rho = \left(4\, \mathfrak{Re}\left(\int_{f_{min}}^{f_{max}}\frac{{\tilde d}(f)\tilde{h}^{*}(f)}{\mathcal{P}(f)}df\right)\right)^{1/2} \, ,
\end{equation}
%%%%---------------------------------------------------------------
where $\rho$ is the amplitude SNR, $d(f)$ represents the data, $h^{*}(f)$ is the complex conjugate of the template, and $\mathcal{P}(f)$ is the power spectra density of the noise. All quantities are Fourier transforms of the time series to the frequency domain. Here, $f_{min}$ is the minimal frequency in the detector sensitivity band and $f_{max}$ is the maximum frequency considered, typically the Nyquist frequency, i.e. half of the sampling frequency.
\textit{Matched filtering}, which finds the template that maximizes the SNR of Eq. (\ref{eq:SNR}), can be shown to be the best possible linear filtering method to retrieve a signal in stationary Gaussian noise. Nonetheless, the noise  of the detectors is neither Gaussian nor stationary, as there are glitches present in the data, which can lead to  a peak in the SNR unrelated to the presence of a  GW signal. Several techniques have been developed to enable the detection and removal of glitches, such as the $\chi^{2}$ test~\cite{Allen:2004gu}. We shall not consider the effect of such glitches in this work.
%%%%---------------------------------------------------------------

The SNR is a measure of the loudness of the signal. The SNR accumulates as the signal enters the sensitivity band of the detector. However, this accumulation happens more slowly for the earlier parts of the signal. When doing early warning, we focus on the early inspiral part of the signal, which is difficult to detect with a short time window. Previous works also intended to implement pre-merger alerts with a matched filtering algorithm, considering only the early inspiral part~\cite{Sachdev:2020lfd}. This makes the search more difficult, as the threshold in SNR must be much lower.

In the following, we use what we shall call the \textit{partial inspiral SNR} (PI SNR) instead of the full optimal SNR.  
The optimal value of the latter can be shown to be the matched filtering of the template with itself \cite{PhysRevD.85.122006},
\begin{equation}\label{eq:optimalSNR}
    \rho_{opt} = \left(4\, \mathfrak{Re}\left(\int_{f_{min}}^{f_{max}}\frac{|\tilde{h}(f)|^2}{\mathcal{P}(f)}df\right)\right)^{1/2} \, .
\end{equation}

\begin{figure}[ht]
\begin{center}
    \includegraphics[width=0.45\textwidth]{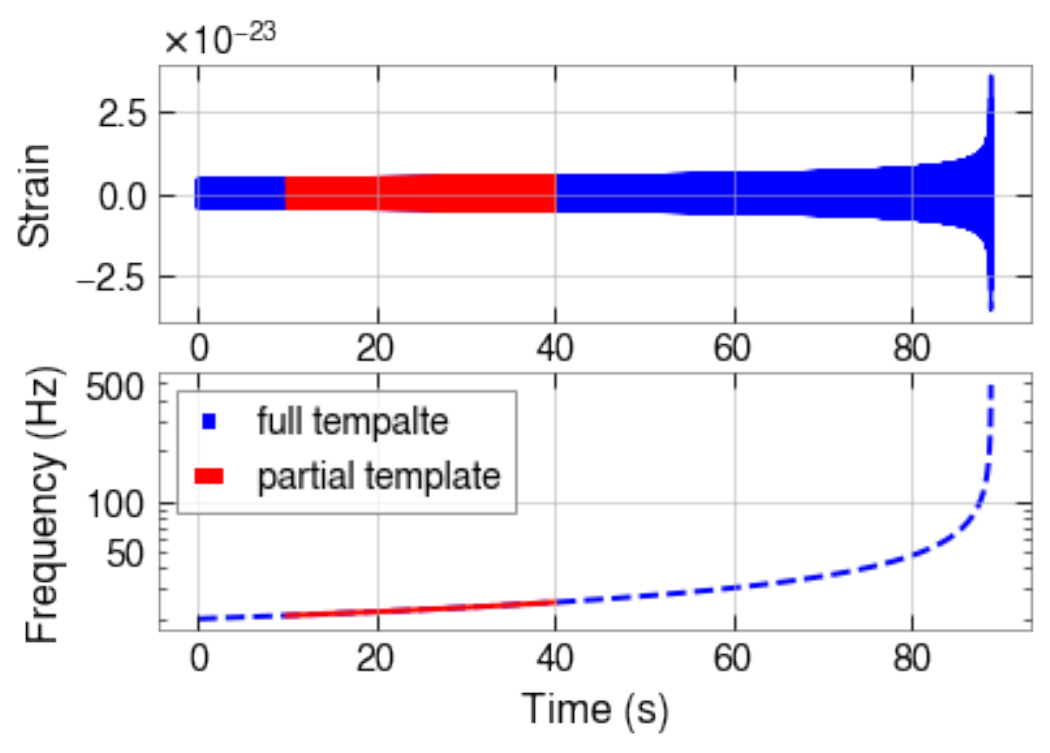}
    \caption{\label{fig:partial_vs_full} The top graph represents a BNS template, where the component masses are of two solar masses, and  the bottom graph shows the evolution in frequency of the template. In both plots the inspiral part under consideration is colored in red.}
\end{center}
\end{figure}

Here, we shall build the PI SNR as the optimal SNR of the considered fraction of template. This is done similarly to the optimal SNR, except that $h$ is replaced by the fraction of the signal under consideration. Typically, the optimal (resp. PI) SNR gives the expected loudness of the signal (resp. part of the signal) in the detector.
In Fig.~\ref{fig:partial_vs_full}, one can see how partial templates compare to the full ones. As the SNR and PI SNR are both inversely proportional to the distance to the BNS system, we can use the latter to change the magnitude of the PI SNR and assess the performance of our CNNs. We shall thus quote a maximum distance at which the CNN can detect the inspiral as a measure of its sensitivity.

%%%%---------------------------------------------------------------

%%%%---------------------------------------------------------------
\subsection{Time-window definition}
%%%%---------------------------------------------------------------
At lowest order in the velocity, the duration of a signal 
in the detectability window of a given detector is~\cite{Samajdar:2021egv}
%%%%---------------------------------------------------------------
\begin{small}
\begin{equation}\label{eq:duration}
\tau \simeq 3\bigg(\frac{1}{\mathcal{M}(M_{\odot})} \bigg)^{\frac{5}{3}}\bigg[ \bigg(\frac{100}{f_{low}(Hz)}\bigg)^{\frac{8}{3}}-\bigg(\frac{100}{f_{high}(Hz)}\bigg)^{\frac{8}{3}}\bigg],
\end{equation}
\end{small}
%%%%---------------------------------------------------------------
where $\mathcal{M}$ is the chirp mass of the system, $f_{low}$ the lowest frequency in the sensitivity window and $f_{high}$ the highest frequency reached by the binary. Eq.~(\ref{eq:duration}) shows that the duration of the signal decreases as a power of $\mathcal{M}$ (and thus of the masses of the components themselves), and of the lowest frequency $f_{low}$ of the detectability band. Since $f_{low}$ will decrease in future detectors, as a consequence the duration of the event will increase. 

Furthermore, at the same order, the SNR can be shown to be \cite{Kastha:2018bar}
%%%%---------------------------------------------------------------
\begin{equation}\label{eq:rho}
{\rm \rho} \simeq \frac{1}{2}\sqrt{\frac{5}{6}}\frac{1}{\pi^{\frac{2}{3}}}\frac{c}{D}\bigg(\frac{G\mathcal{M}}{c^{3}} \bigg)^{\frac{5}{6}}\sqrt{I(M)}g(\theta, \phi, \psi, \iota).
\end{equation}
%%%%---------------------------------------------------------------
In this expression, $c$ is the speed of light, $D$ is the luminosity distance to the BNS system, $G$ is the gravitational constant, $\mathcal{M}$ is the chirp mass, $I(M)$ is the frequency integral, which depends on the PSD and the frequency range of the detector, $M$ the total mass of the binary, and $g(\theta, \phi, \psi, \iota)$ is a function that depends on the sky position, through the antenna pattern of the detectors \cite{Schutz:2011tw}. 
From Eq. (\ref{eq:duration}) and Eq. (\ref{eq:rho}), when all the parameters except the masses are fixed, the CBC signal becomes longer and weaker as the masses of the progenitors are reduced. Hence, there is a strong need to find a balanced window size, large enough for the inspiral to be detectable, and short enough for the detection to be fast. A summary of the durations of the \textit{time windows} (TW) used for each category of CBC can be seen in Table~\ref{tab:inputs}. Indeed, the TW needed depends on the duration of the signal, which is related to the component masses. The alert time before the merger is defined as the difference between the duration of the signal and the TW. Therefore,  each category represents a typical component mass range as well as a different input size. We then train one adapted CNN for each category and fine-tune it for the considered situation. In this way we can reduce our classification problem to three smaller binary-classification tasks. For each task we have only two classes:  a \textit{noise class} with only noise, and an \textit{event class} with noise and inspiral. Note that the classes are balanced, i.e. half of our data are noise-class instances and the other half are event-class instances.

%%%%---------------------------------------------------------------
\begin{table}
\begin{center}
\begin{tabular}{|c|c|c|c|}
  \hline
  Object & light \texttt{BNS} & intermediate \texttt{BNS} & heavy \texttt{BNS}  \\
  \hline
  \hline
  Mass range (\(M_\odot\)) & 1.4 - 1.8 & 1.8 - 2.4 & 2.4 - 3 \\\hline
  Freq. cutoff (Hz) & 20 & 20 & 20 \\\hline
  Duration (s) & 160 - 100 & 100 - 65 & 65 - 45  \\\hline
  Time window (s) & 80 & 50 & 30 \\\hline
  Fraction of signal & 0.5 - 0.8 & 0.5 - 0.77 & 0.46 - 0.66\\\hline
  Early alert  & 80 - 20 & 50 - 15 & 35 - 15   \\
  before merger (s) & & &  \\
  \hline
\end{tabular}
\end{center}
\caption{Summary of the type of CBC mergers that each CNN is trained for. Due to the difference in duration of the signals for different masses, the TW considered are different from one network to the other. The TW used here will be optimized in a later work.}
\label{tab:inputs}
\end{table}

%%%%---------------------------------------------------------------
\subsection{Data generation}
\begin{figure}[ht]
\begin{center}
    \includegraphics[scale=0.7]{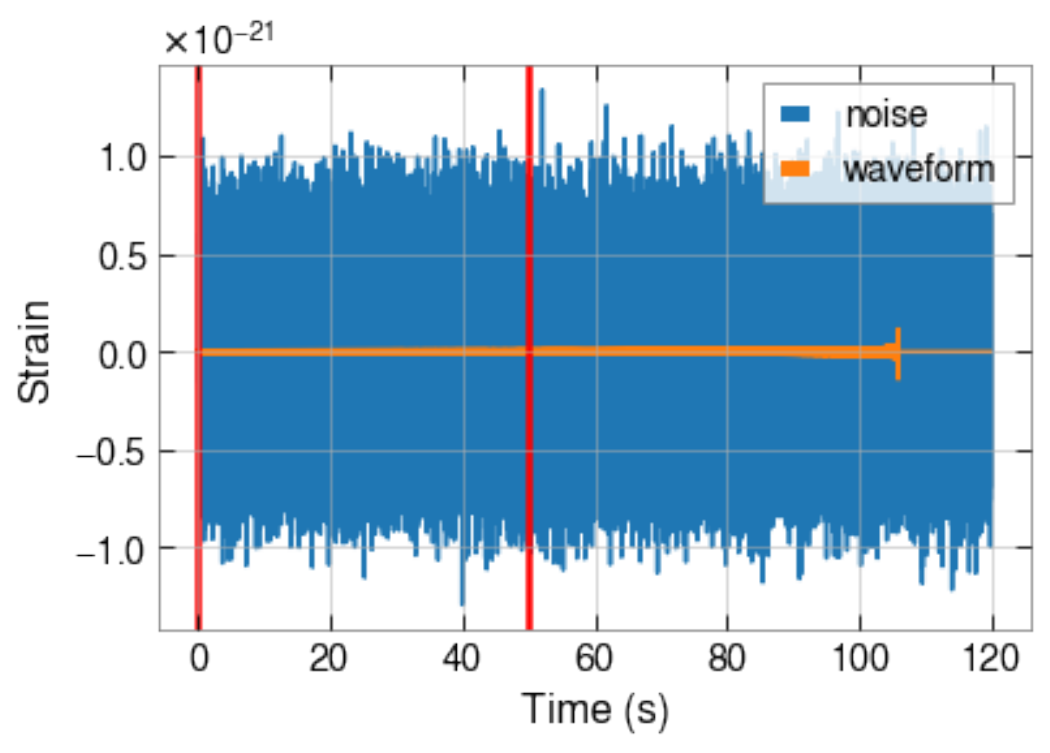}
    \caption{\label{fig:SigAndNoise} Representation of the noise and the injected data. The CBC signal corresponds to a BNS where both component masses are $1.8 M_{\odot}$ and the binary is placed at a distance of $100$ Mpc. When training and testing the network, we do not pass this full frame to the network, but only the first 50 s (between the two red line) which is the chosen duration for the inspiral.}
\end{center}
\end{figure}

We generate two 1D whitened time series sets for training and testing with the \textit{PyCBC} package \cite{Biwer:2018osg}. The steps to generate our data are the following:
%%%%---------------------------------------------------------------
\begin{itemize}
    \item We sample 120 s of Gaussian noise. This noise is coloured via a PSD that corresponds to the aLIGO design sensitivity. We take the minimal frequency for the data to be 15 Hz.
    \item We generate a GW signal and inject it into the data. Here, we considered only non-spinning binaries and the template is \textit{SpinTaylorT4} ~\cite{Buonanno:2002fy}. We project the waveform onto the detector using the antenna pattern\footnote{The antenna patter is a function representing the response of the detector to the sky position, see~\cite{schutz2011networks}. We chose here $\theta = \phi = 60^{\rm o}, \psi=0^{\rm o}$ for the angle in the antenna pattern formula.}. The signal is then injected into the noise. A representation of the noise and the injected waveform is shown in Fig.~\ref{fig:SigAndNoise}. 
    As the noise is generated to 15 Hz, we chose to generate the waveform at 20 Hz to be sure to avoid any border effect when the whitening is performed.
    \item We select the appropriate duration of the window that we want to analyze (for example 50 s for the intermediate BNSs), which corresponds to a chosen duration of the inspiral. This signal is then whitened and the PI SNR is computed for this part of the signal.
\end{itemize}
%%%%---------------------------------------------------------------
%%%%---------------------------------------------------------------
\begin{figure}[ht]
\begin{center}
    \includegraphics[width=0.55\textwidth]{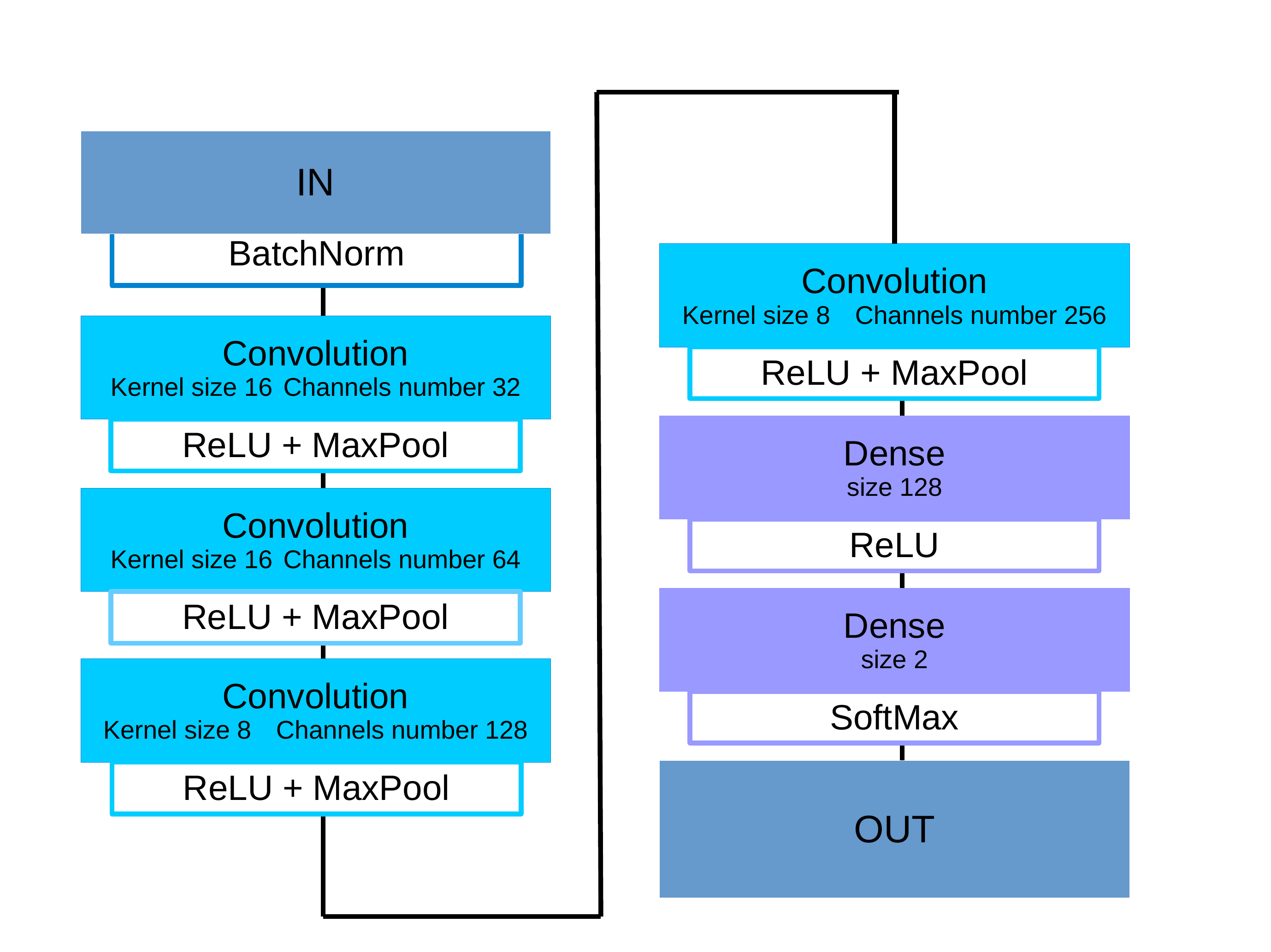}
    \caption{\label{fig:config} Best-performing architectures for our 4 different CNNs. All the MaxPool layers have a kernel size and a stride of 4.}
\end{center}
\end{figure}

\section{Methodology and Convolutional Neural Networks}
\label{sec:method}
\subsection{Training and testing of our neural networks}

As was previously mentioned, the data sets are balanced to improve the performance of the networks: 
$50 \%$ of the data belongs to the \textit{event class}  and the other $50 \%$ to the \textit{noise class}. Each training set contains 8000 TW, built as described in Sec.~\ref{sec:PartaiSNR}. From this set we take $80 \%$ of the data for training and $20 \%$ for validation. In this way, the network optimizes its learnable parameters with the training set, and we test its performance with the unseen validation set. This is a standard practice in DL, as we can monitor the generalisation ability of the network at each epoch.  
Finally, we test it with a balanced set consisting of $4000$ TW.

%%%%---------------------------------------------------------------

The performance of our networks is assessed on the basis of the true alarm probability (TAP) for a given false alarm probability (FAP). These are defined as 
\begin{equation}\label{eq:TAP}
TAP = \frac{TP}{TP+FN} \qquad FAP = \frac{FP}{TN + FP},
\end{equation}
where TP, FN, TN, FP are respectively the number of true positives, false negatives, true negatives and false positives, according to the standard \textit{confusion matrix}. The TAP
is given by the number of events that are correctly classified as \textit{event class} over the total number of frames that actually contain one. The FAP is given by the number of TW that are correctly classified as \textit{noise class} divided by the total number of frames that do not contain an inspiral. By fixing this value, we ensure that there is a reasonable precision, while by maximizing the TAP we ensure a large sensitivity. It is important to note that FAP is related to the decision threshold.

%%%%---------------------------------------------------------------

For this work, we chose the FAP to be $1\%$. This value can be seen as high compared to the values that are used in the traditional matched filtering. Nonetheless, let us stress that this work is a proof of concept to  be used later in a full pipeline. In the future, a DL pipeline will use coincidence between various detectors, which will improve the performance of the network and significantly lower the FAP.  
As explained in Sec.~\ref{sec:PartaiSNR}, the performance of the networks is a function  of the distance $D$. Thus, we  divide the test set in batches according to this parameter. A description of the training strategy can be found in~\cite{baltus2021convolutional}.

%%%%--------------------------------------------------------------
\subsection{Architecture of the CNN}
%%%%---------------------------------------------------------------

For the development of our networks, including the model definition, the training and the validation phases, we have used the PyTorch framework \cite{NEURIPS2019_9015}. It is a standard practice to employ Adam as an optimizer, but it might fall into bad local minima. Thus, we use Adamax instead, because it has shown better performance. This optimizer is a variant of Adam based on the infinity norm \cite{kingma2019method}.
We use a cross-entropy loss function and a softmax activation function for the last layer. In such a way, the output of the CNN will be a tuple with the probabilities that a certain window belongs to the \textit{noise class} and the \textit{event class}. As we mentioned before, we implement short CNNs ($\sim 10$ layers at most)
to see if reasonable performances can be obtained. After some architecture optimizations, it was found that stacking more than 4 layers did not improve our results. Thus, we present the best-performing architecture in Fig.~\ref{fig:config}.  We present the performance of this network for the different categories in  Sec.(\ref{sec:Discussion}). 

%%%%---------------------------------------------------------------

\begin{figure}[t]
\begin{center}
    \includegraphics[keepaspectratio, width=0.4\textwidth]{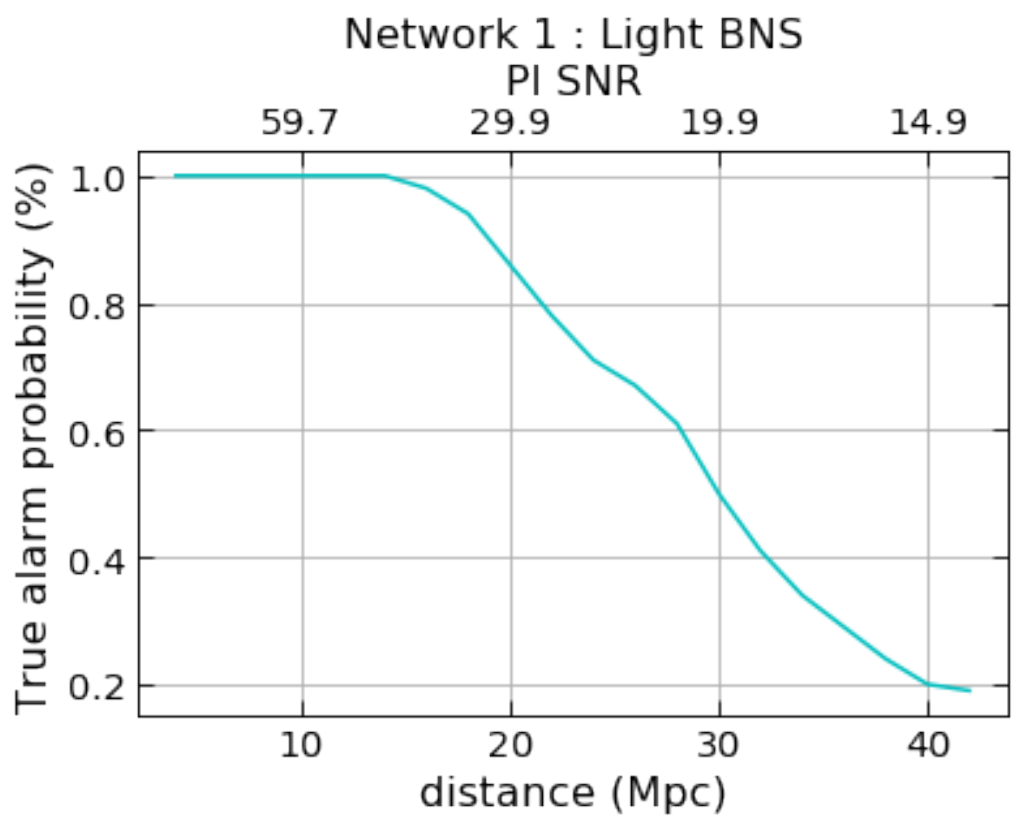}
    \includegraphics[keepaspectratio, width=0.4\textwidth]{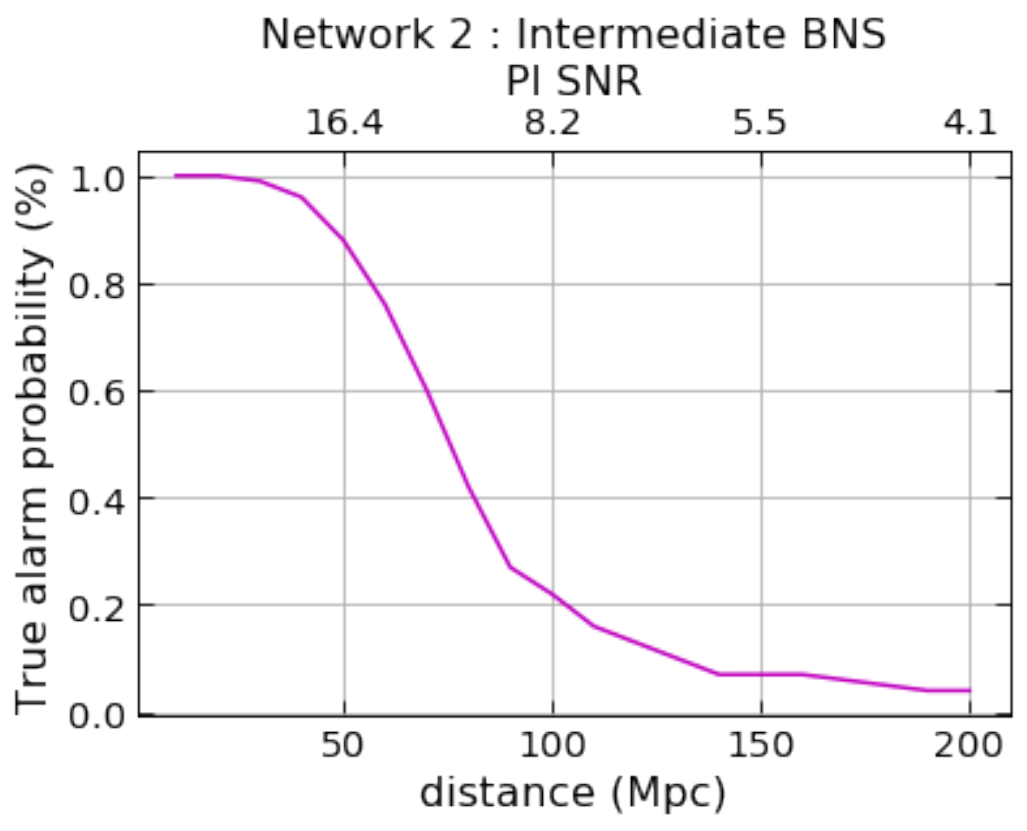}
    \includegraphics[keepaspectratio, width=0.4\textwidth]{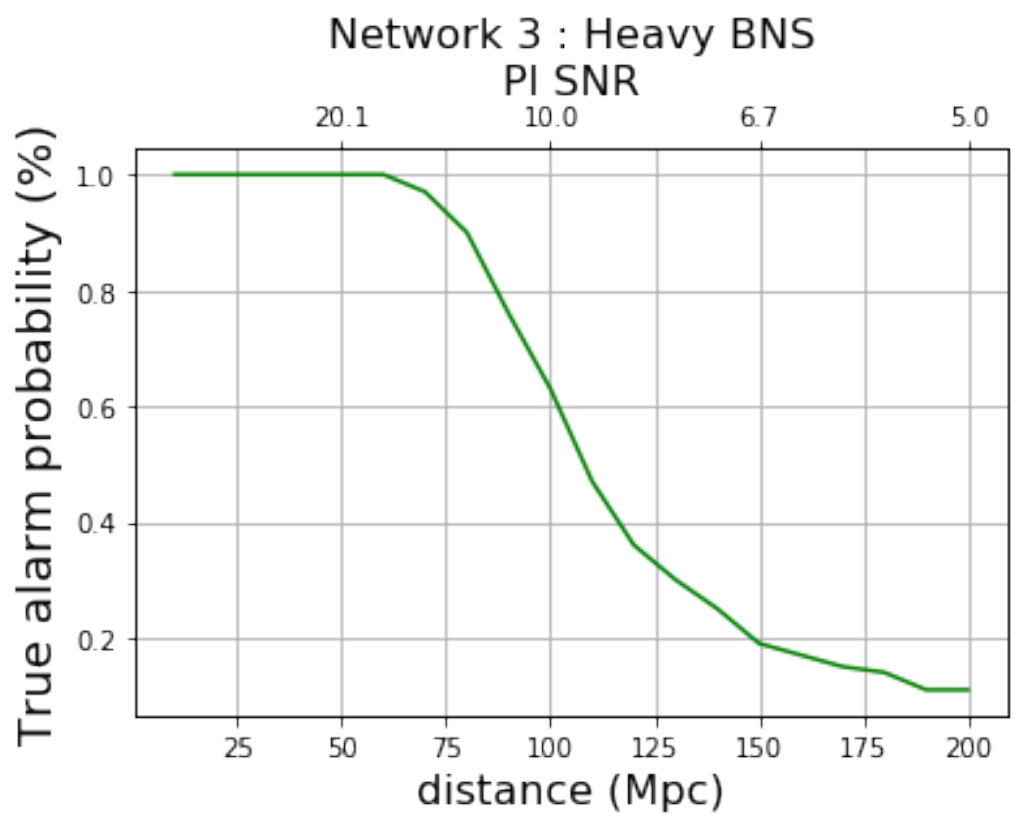}
    \caption{\label{fig:SumResults} 
    TAP as a function of the distance and PI SNR for a fixed FAP at 0.01. These curves correspond to 200 instances at every fixed distance.
    }
\end{center}
\end{figure}
\section{Results and discussion}\label{sec:Discussion}

Our network has been trained and tested on each of the categories described in section~\ref{sec:PartaiSNR}.  A summary of the evolution of the TAP as a function of the distance for each case is shown in Fig~\ref{fig:SumResults}. We can see that the network is able to detect events farther away when the masses are larger. 
This dependence is expected, and a similar one has been observed in matched filtering and in previous studies made with CNN-based neural networks~\cite{George:2016hay}. 
This is due to the fact that the amplitude of the signal is proportional to $\mathcal{M}^{\frac{5}{3}}$  \cite{postnov2006evolution}, while the amplitude of the noise is constant, as it is described by the PSD (see Sec. II). As a consequence, for larger masses, events can be detected at greater distances.
 
We can now focus on each category individually. First, for the light BNS systems, we have a TAP remaining close to 1 up to $15$ Mpc and still at 0.2 for $40$ Mpc, that corresponds to the distance of   GW170817 \cite{abbott2017gw170817}. 
As a consequence, we see that there is a fair chance that our network will detect such realistic events in advance once the interferometers are at design sensitivity. 

For the intermediate BNS inspirals, the corresponding network is able to detect the signal above $100$ Mpc. The curve is slightly better for the heavy BNS. Another characteristic worth looking is the PI SNR evolution for the templates. In our tests, the TAP starts to decrease when the PI SNR reaches values between 15 and 20, except for the light BNS system where the results are slightly worse. 

A direct comparison with matched filtering is non-trivial. As mentioned earlier, most of the matched filtering searches use the full template and the optimal SNR. In our search, we used only the early inspiral part of the signal and the PI SNR. Nevertheless, a comparison of our method with ~\cite{Sachdev:2020lfd} on a realistic BNS population can be found in~\cite{baltus2021convolutional}.

We also tested a network trained on a particular type of objects (e.g. heavy BNS) on another type (e.g. median BNS). To do so, we changed the size of the TW considered for the latter object type to the size used for the former one (e.g. 30 seconds). The performances of our CNN degraded considerably, dropping frorm $\sim 65\%$ on the heavy BNS class to $\sim 5\%$ for the median BNS class.

\begin{figure}[ht]
\begin{center}
    \includegraphics[keepaspectratio, width=0.45\textwidth]{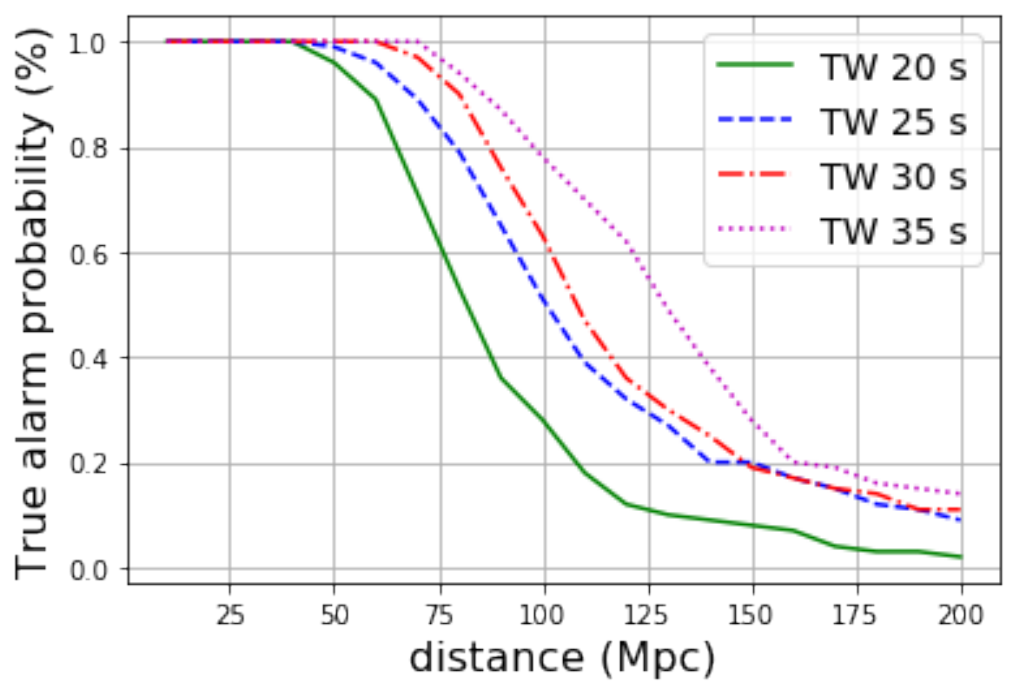}
    \caption{\label{fig:TW_evol} The performance of the network trained on the heavy BNS category with different TW. For all of those curves the TAP is defined for a fixed FAP of 0.01.}
\end{center}
\end{figure}

The TW chosen for a particular category of objects has a significant impact on the efficiency of the network. In Fig.~\ref{fig:TW_evol}, we can observe how the performance of a network trained on heavy BNS increases as the TW is enlarged. This is expected, because when increasing the TW we consider a  wider amplitude and a larger fraction of the signal making it easier to detect. Based on this observation, we could train the network for a given category on several TW, which would help detect the loudest events faster while enabling larger accumulations for fainter signals.

Our method is fast and does not require a lot of resources, contrarily to the traditional techniques. Indeed, the computation time needed to perform matched filtering with only one template on a 50 second frame is 0.05 s on an Intel(R) Core(TM) i7-8650U CPU. This represents a lower bond because matched filtering must be applied on the whole template bank. We note that our method only takes 0.005 s to analyze the same frame with an Nvidia GeForce RTX 2070 SUPER GPU.

\section{Conclusion}\label{sec:Conclusions}
In this work, we have shown that small CNNs can be the building blocks of a larger and fast early alert pipeline. With the lengthening and multiplication of signals, the development of such a pipeline may become crucial for MMA, since CNNs are less computationally expensive.

We have performed a first optimization, 
for the heavy BNS systems, as shown in Fig.~\ref{fig:TW_evol}. Because the best configuration of the CNN (kernel sizes and architecture) might depend on the CBC type under consideration and the size of the TW.
Further optimization is needed to extract the best performances of our set-up.

In a future research, we  plan to explore other architectures in order to improve the performance of the algorithm. The training method can also be enhanced using \emph{curriculum learning} and we could include Reinforcement Learning to automatize this process. Another crucial future development will be the extension of these methods to a network of interferometers, crucial for sky localisation. From the analyses based on matched filtering, we know that the consideration of several data streams and the requirement of coincident detections lowers considerably the FAP, behaving as $ 0.01 N_{det}$, where $N_{det}$ is the number of detectors.

Therefore, the detection probabilities in this paper are really lower bounds, that will be improved in the near future through fine tuning of hyper-parameters, the introduction of specific tools for machine learning and coincidence among numerous detectors.
These ideas combined will be transformed into a unique pipeline.

Machine Learning techniques have shown great performance in different researches, and this proof of concept work is not an exception. Hence, we will continue with  this line of research to address future challenges of GW data analysis.

\section*{Acknowledgments}
The authors thank Michal Bejger and Andrew Miller for useful comments.
G.B. is supported by a FRIA grant from the Fonds de la Recherche Scientifique-FNRS, Belgium.
J.R.C. acknowledges the support of the Fonds de la Recherche Scientifique-FNRS, Belgium, under grant No. 4.4501.19.
M.L., S.C, and J.J  are supported by the research program 
of the Netherlands Organisation for Scientific Research (NWO).
The authors are grateful for computational resources provided by the 
LIGO Laboratory and supported by the National Science Foundation Grants No. PHY-0757058 and No. PHY-0823459.

\bibliography{bare_conf.bib}{}
\bibliographystyle{IEEEtran}
\end{document}